\documentclass[fleqn,usenatbib]{mnras}

\usepackage{newtxtext,newtxmath}

\usepackage[T1]{fontenc}
\usepackage{ae,aecompl}

\usepackage{graphicx}	
\usepackage{amsmath}	
\usepackage{bm}
\usepackage{ulem}
\usepackage{dblfloatfix}

\usepackage{scalerel,tikz}
\usetikzlibrary{svg.path}
\definecolor{orcidlogocol}{HTML}{A6CE39}
\tikzset{orcidlogo/.pic={
 \fill[orcidlogocol] svg{M256,128c0,70.7-57.3,128-128,128C57.3,256,0,198.7,0,128C0,57.3,57.3,0,128,0C198.7,0,256,57.3,256,128z};
 \fill[white] svg{M86.3,186.2H70.9V79.1h15.4v48.4V186.2z}
 svg{M108.9,79.1h41.6c39.6,0,57,28.3,57,53.6c0,27.5-21.5,53.6-56.8,53.6h-41.8V79.1z M124.3,172.4h24.5c34.9,0,42.9-26.5,42.9-39.7c0-21.5-13.7-39.7-43.7-39.7h-23.7V172.4z}
 svg{M88.7,56.8c0,5.5-4.5,10.1-10.1,10.1c-5.6,0-10.1-4.6-10.1-10.1c0-5.6,4.5-10.1,10.1-10.1C84.2,46.7,88.7,51.3,88.7,56.8z};
}}
\newcommand\orcidicon[1]{\href{https://orcid.org/#1}{\mbox{\scalerel*{
\begin{tikzpicture}[yscale=-1,transform shape]
\pic{orcidlogo};
\end{tikzpicture}
}{|}}}}

\title[Impact of filaments in Auriga]{The impact of filaments on dwarf galaxy properties in the Auriga simulations}

\author[H. Zheng et al.]{
Haonan Zheng\orcidicon{0000-0002-1665-5138} $^{1,2}$\thanks{Email: hnzheng@nao.cas.cn}, 
Shihong Liao$^{1,3}$\thanks{Email: shliao@nao.cas.cn}, 
Jia Hu$^{1,2}$, 
Liang Gao$^{1,2,4}$, \newauthor\ 
Robert J. J. Grand$^{5,6,7}$, 
Qing Gu$^{1,2}$, 
Qi Guo$^{1,2}$
\\
\\
  $^{1}$Key Laboratory for Computational Astrophysics, National Astronomical Observatories, Chinese Academy of Sciences, Beijing 100012, China\\
  $^{2}$University of Chinese Academy of Sciences, 19 A Yuquan Rd, Shijingshan District, Beijing 100049, China\\
  $^{3}$Department of Physics, University of Helsinki, Gustaf H\"{a}llstr\"{o}min katu 2, FI-00014 Helsinki, Finland\\
  $^{4}$Institute for Computational Cosmology, Department of Physics, University of Durham, South Road, Durham, DH1 3LE, UK\\
  $^{5}$Max-Planck-Institut f\"{u}r Astrophysik, Karl-Schwarzschild-Str. 1, 85748 Garching, Germany\\
  $^{6}$Instituto de Astrof\'isica de Canarias, Calle V\'ia L\'actea s/n, E-38205 La Laguna, Tenerife, Spain\\
  $^{7}$Departamento de Astrof\'isica, Universidad de La Laguna, Av. del Astrof\'isico Francisco S\'anchez s/n, E-38206, La Laguna, Tenerife, Spain\\
}

\date{Accepted XXX. Received YYY; in original form ZZZ}

\pubyear{2021}

\defcitealias{Liao&Gao2019}{LG19}

\begin{document}
\label{firstpage}
\pagerange{\pageref{firstpage}--\pageref{lastpage}}
\maketitle

\begin{abstract}

With a hydrodynamical simulation using a simple galaxy formation model without taking into account feedback, our previous work has shown that dense and massive filaments at high redshift can provide potential wells to trap and compress gas, and hence affect galaxy formation in their resident low-mass haloes. In this paper, we make use of the Auriga simulations, a suite of high-resolution zoom-in hydrodynamical simulations of Milky Way-like galaxies, to study whether the conclusion still holds in the simulations with a sophisticated galaxy formation model. In agreement with the results of our previous work, we find that, comparing to their counterparts with similar halo masses in field, dwarf galaxies residing in filaments tend to have higher baryonic and stellar fractions. At the fixed parent halo mass, the filament dwarfs tend to have {slightly} higher star formation rates than those of field ones. {But overall we do not find a clear difference in galaxy $g-r$ colours between the filament and field populations.} We also show that at high redshifts, {the gas components} in dwarf galaxies tend to have their spins aligned with the filaments in which they reside. Our results support a picture in which massive filaments at high redshift assist gas accretion and enhance star formation in their resident dwarf sized dark matter haloes.   

\end{abstract}




\section{Introduction}
\label{sec:intro}

In the standard galaxy formation framework \citep[e.g.][]{White&Rees1978, White&Frenk1991}, dark matter haloes provide potential wells to compress gas and then the subsequent physics, e.g. radiative cooling, star formation, stellar and black hole feedback and so on, proceed to form galaxies. Similarly to dark matter haloes, dense and massive filaments at high redshift can also provide deep potential well to trap and compress gas and provide sites for galaxy formation.

The impact of filaments on galaxy formation is particularly prominent in warm dark matter (WDM) models: \citet{Gao2007} show that the first stars can directly form in the potential wells of massive filaments in WDM models; \citet{Gao2015} further show with hydrodynamical simulations that smooth and dense filaments rather than haloes dominate star formation at $z \ga 6$ in WDM models. In the cold dark matter (CDM) paradigm, as shown by \citet{Keres2005} and many other follow-up works, CDM filaments can direct gas flows to the centres of massive galaxies \citep[see also e.g.][]{Dekel2006,Ocvirk2008,Brooks2009,Dekel2009,Keres2009}. Recently, with a hydrodynamical version of the zoom-in Aquarius simulation \citep{Springel2008}, \citet{Liao&Gao2019} \citepalias[hereafter][]{Liao&Gao2019} further show that, apart from feeding massive central galaxies, high redshift filaments indeed compress gas, and so assist gas cooling and enhance star formation in their resident low mass dark matter haloes.

In \citetalias{Liao&Gao2019}, in order to investigate gas accretion and cooling in filaments and their low-mass dark matter haloes, as a first step, the authors neglected the stellar feedback effects in their simulations. In this work, we use the Auriga simulations \citep{Grand2017}, a series of high-resolution runs with more realistic galaxy formation models (especially including supernovae and black hole feedback processes), to study the impact of filaments on galaxy formation in their resident dwarf dark matter haloes.

This paper is organized as follows. Section \ref{sec:methods} describes the details of simulations and our filament identifications method. Our results and conclusions are presented in Sections \ref{sec:results} and \ref{sec:conclusion} respectively. Resolution studies are given in an Appendix.

\begin{figure*}
    \centering
	\includegraphics[width=2.0\columnwidth]{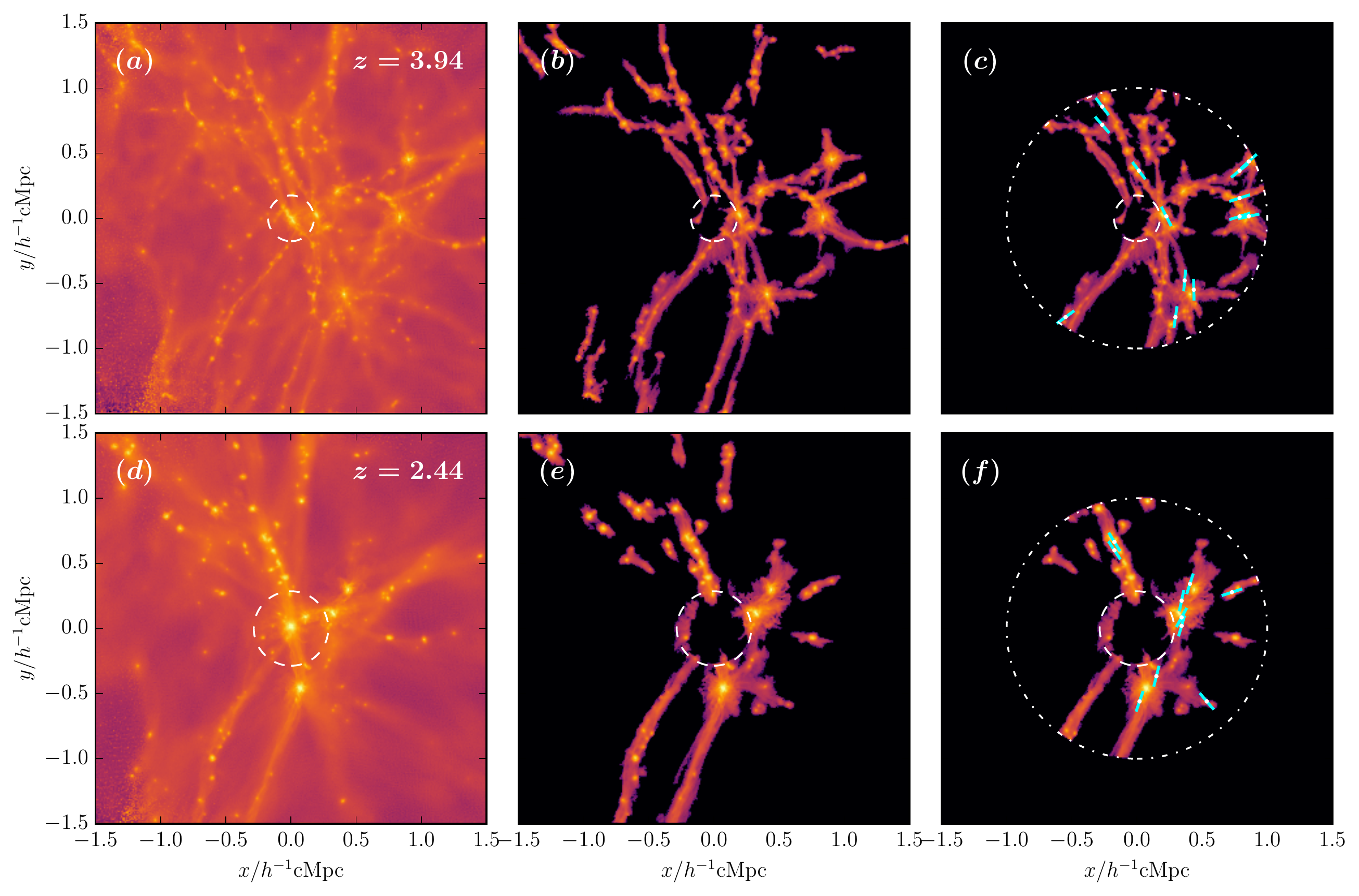}
	
	\caption{Panel (a): $xy$-projected cloud-in-cell baryonic density field of the Au-27 simulation at $z=3.94$. The white dashed circles in all panels marks the radius $2R_{\rm 200, MW}$ centred on the main halo. Panel (b): filaments identified with the method described in \citetalias{Liao&Gao2019}.
	Panel (c): short cyan lines overplotted on the density field show the projected directions of the filament at random galaxies. 
	The dash-dotted circle shows the radius of $1.0 ~  h^{-1}\mathrm{cMpc}$ within which our galaxy sample is chosen. 
	Panels (d)-(f): similar to panels (a)-(c), but for redshift of $z=2.44$.}
	
	\label{fig:fig1}
\end{figure*}

\section{The Simulations and filament identification}
\label{sec:methods} 
We use the hydrodynamical zoom-in simulations of the Auriga project \citep{Grand2017}, which consist of 30 isolated Milky Way-sized galaxies and their surroundings with the \textsc{\MakeLowercase{AREPO}} code \citep{Springel2010}. Each Auriga simulation is labelled as `Au-$X$' with $X$ ranging from 1 to 30. 
The cosmological parameters correspond to the Planck results: $\Omega_{\rm{m}} = 0.307$, $\Omega_{\Lambda} = 0.693$, $\Omega_{\rm{b}} = 0.048$, and $h = 0.6777$ \citep{Planck2014p16}.
For the level-4 resolution simulations that we use in this analysis, the initial masses of baryon and high-resolution dark matter particles are $m_{\rm b} \sim 5\times 10^{4} ~ \mathrm{M}_{\sun}$ and $m_{\rm DM} \sim 3\times 10^{5} ~ \mathrm{M}_{\sun}$ respectively. The gravitational softening length, $\epsilon$, for star and high-resolution dark matter particles is set as $500 ~ h^{-1} \mathrm{cpc}$ before $z=1$, and fixed at $369 ~  \mathrm{pc}$ later. As for gas cells, the softening length is adjusted according to their mean radius, from $500 ~ h^{-1} \mathrm{cpc}$ ($369 ~ \mathrm{pc}$ after $z=1$) to $1.85 ~ \mathrm{kpc}$. Haloes and subhaloes are identified with the friends-of-friends \citep[FOF,][]{Davis1985} and \textsc{\MakeLowercase{SUBFIND}} algorithm \citep{Springel2001} respectively. We refer the reader to \citet{Grand2017} for a detailed description of the Auriga simulations.

To identify filaments in each Auriga simulation, we use the method proposed in \citetalias{Liao&Gao2019}, which is based on the baryonic density field and the Hoshen-Kopelman algorithm \citep{Hoshen&Kopelman1976}.
In this method, we consider all baryonic particles/cells (i.e. gas and stars) in the zoom-in simulation region. Note that the zoom-in region in different Auriga runs has different shapes and spatial extents, and we have chosen a fixed $3 ~ h^{-1} \mathrm{cMpc}$ cubic region centring the Milky Way-like galaxy centre\footnote{The centre of a galaxy is defined as the position of the particle with the minimum gravitational potential energy in the FOF group.} for all 30 Auriga runs. Then we compute the baryonic density field of this cubic region on $512^3$ grid cells using the clould-in-cell (CIC) method. See Panels (a) and (d) of Fig. \ref{fig:fig1} for examples of the projected baryonic density fields at two redshifts from the Au-27 simulation. We further exclude the central region within twice of the virial radius of the main halo (the dashed circles in Fig. \ref{fig:fig1}) as well as the cells whose overdensity is lower than a given threshold $\Delta_{\mathrm{th}}$, link neighbouring cells as structures using the Hoshen-Kopelman algorithm, and define those structures with sizes (i.e. the number of containing grid cells) larger than a given size threshold $S_{\mathrm{th}}$ as `filaments'. The rest of the region outside the excluded central sphere is defined as `field'. Like the overdensity parameter in spherical overdensity halo finder, $\Delta_{\mathrm{th}}$ is a free parameter to control the overdensity and boundary of a filament, and $S_{\mathrm{th}}$ is used to exclude small isolated high-density regions (which are usually related to isolated large galaxies) from filament classification. These two free parameters regulate how prominent the detected filaments are, with higher values for them, the detected filaments will be more overdensed and have larger sizes.
In this work, the two threshold parameters are set to $\Delta_{\rm th} = 3.0$ and $S_{\rm th} = 3000${, which return good filament structures according to human eyes' judgements; see Panels (b) and (e) of Fig. \ref{fig:fig1} for examples}. Following \citetalias{Liao&Gao2019}, we also only consider high-redshift dense filaments at $z\sim 4$ and $\sim 2.5$ in this study, as low redshift filaments have less impact on galaxy formation \citep{Gao2015}. 
Note, we have varied the values of $\Delta_{\rm th}$ and $S_{\rm th}$ and confirmed that the results presented in the following sections are not sensitive to the choices of these parameter values. {For example, we have tried with $(\Delta_{\rm th}, S_{\rm th}) = (2.0, 3000), ~(4.0, 3000),~ (3.0, 2000), ~(3.0, 4000),$ and $(3.5, 1000)$, the quantitative difference of the baryonic fraction - virial mass relations (which will be discussed in the next section) is just a few per cent.}

We have also varied the number of grid cells when computing the CIC density field in the $3~h^{-1}{\rm cMpc}$ cubic region, and made sure that our results do not sensitively depend on the number of grid cells. For example, we have tested with $256^3$  grid cells, and with $\Delta_{\rm th} = 3.0$ and $S_{\rm th}=375$, the baryonic fraction - virial mass relation only differ from our fiducial results at a level of a few per cent.

{In this study, a galaxy consists of all types of particles/cells within the virial radius\footnote{The virial radius of a galaxy is defined as the radius within which the mean mass density is 200 times the critical density of the universe.}, $R_{200}$, of the corresponding FOF group.} If the centre of a galaxy is located in a classified filament (field) environment, then this galaxy is called as a filament (field) galaxy. Note that in order to avoid the influence from the main galaxies and contamination by low-resolution particles from the outer parts of the simulation volume, we only use galaxies whose comoving distance from the main galaxy centre,  $d$, satisfies $2R_{\rm 200,\ MW} < d < 1.0 ~ h^{-1} \mathrm{cMpc}$, the $R_{\rm 200,\ MW}$ denotes the comoving virial radius of the main central galaxy. 
In our parent galaxy sample, we only consider galaxies with virial masses (i.e. the total mass within $R_{200}$) $M_{200} \geq 10^8~h^{-1}{\rm M}_\odot$, which contain at least 350 dark matter particles.
In total, at $z = 3.94\ (2.44)$, we have $4192\ (2736)$ filament galaxies and $11036\ (13464)$ field galaxies in our {parent} sample from all thirty Auriga simulations. {Note that the resolution requirement of $\geq 350$ dark matter particles for the parent sample is similar to those adopted in previous literature when studying halo baryonic fractions \citep[e.g.][]{Crain2007,Okamoto2008} and stellar fractions \citep[e.g.][]{Puchwein2013}. As detailed in the following sections, we will place additional resolution requirements on the baryonic particle number when computing the star formation rates, galaxy colours, and galaxy spins.}

Note that apart from the aforementioned thirty Level 4 resolution runs, the Auriga project has additional six Level 3 runs with $\sim 8$ times higher resolution (i.e. Au-6, Au-16, Au-21, Au-23, Au-24, and Au-27). We have performed a resolution convergence study with six Auriga simulations from both the Level 3 and Level 4 runs, and confirm that the results presented in the following sections do not suffer from numerical resolution issues (see Appendix \ref{ap:resolution} for details). In order to have better statistics, in the following we mainly show results from the thirty Level 4 runs.

\begin{figure*}
    \centering
	\includegraphics[width=2.0\columnwidth]{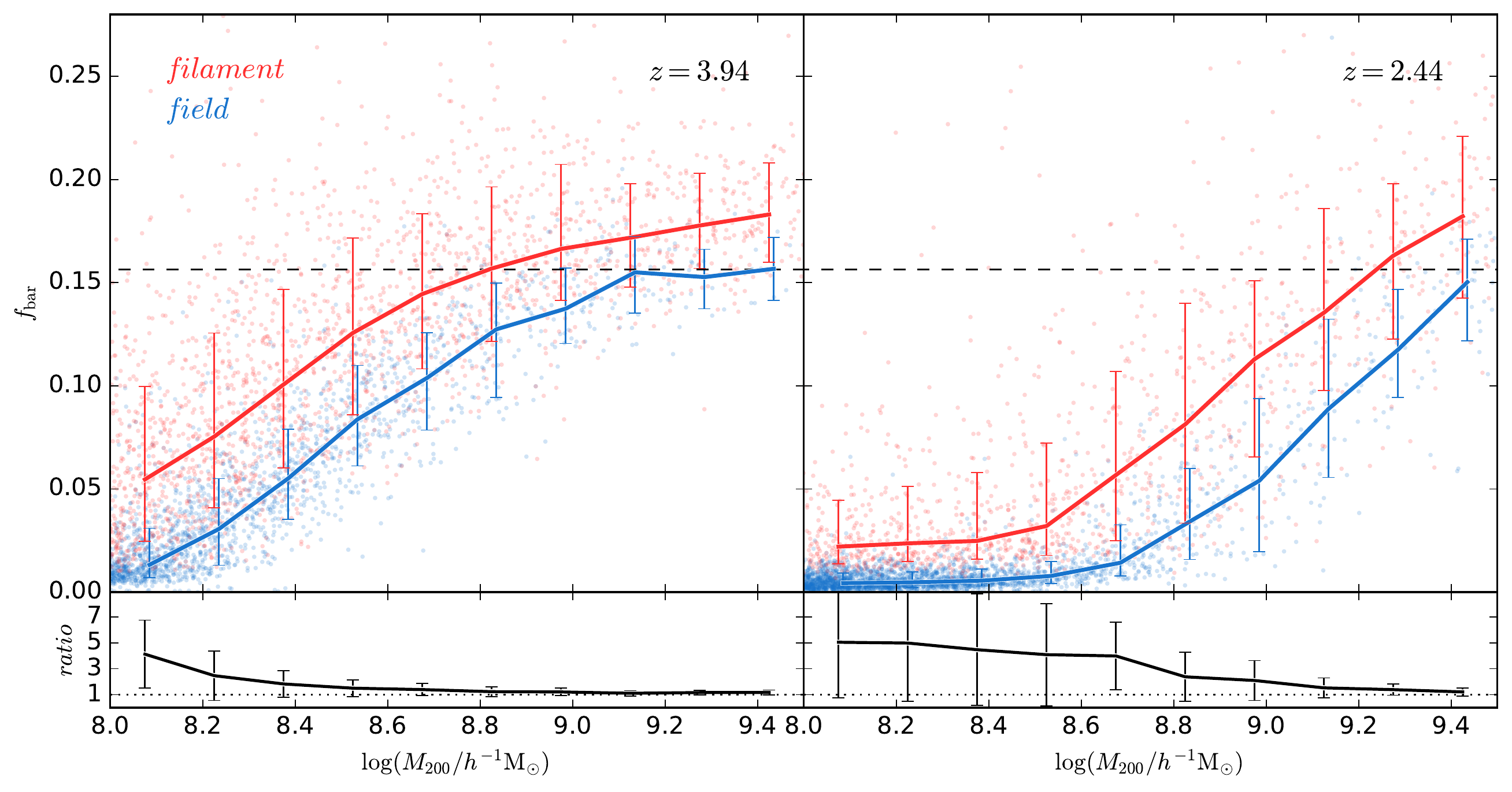}
	
	\caption{Relation between baryonic fraction, $f_\mathrm{\,bar}$, and galaxy virial mass, $M_\mathrm{200}$, at $z=3.94$ (left) and $2.44$ (right). In the upper panels, the galaxies in filaments and field are shown with red and blue colours, respectively. The solid lines show the median baryonic fractions, while the error bars show the 16th and 84th percentiles of the baryonic fraction in each mass bin. The horizontal dashed line indicates the cosmic baryonic fraction ($\mathrm{\Omega_b} / \mathrm{\Omega_m}$ = 0.156). In the bottom panels, the black lines show the ratios between the median baryonic fractions in filament and field galaxies in different mass bins, and the error is derived by error propagation after symmetrizing the asymmetric errors plotted in the upper panels; see the main text for details. The dotted horizontal line in the bottom panels marks the ratio of 1. Note, we only plot galaxies with $M_{200} \leq 10^{9.5}~h^{-1}{\rm M}_\odot$, as there are very few field galaxies with masses $M_{200} \ga 10^{9.5}~h^{-1}{\rm M}_\odot$ in the Auriga simulations. }
	
	\label{fig:fig2}
\end{figure*}

\begin{figure*}
    \centering
	\includegraphics[width=2.03\columnwidth]{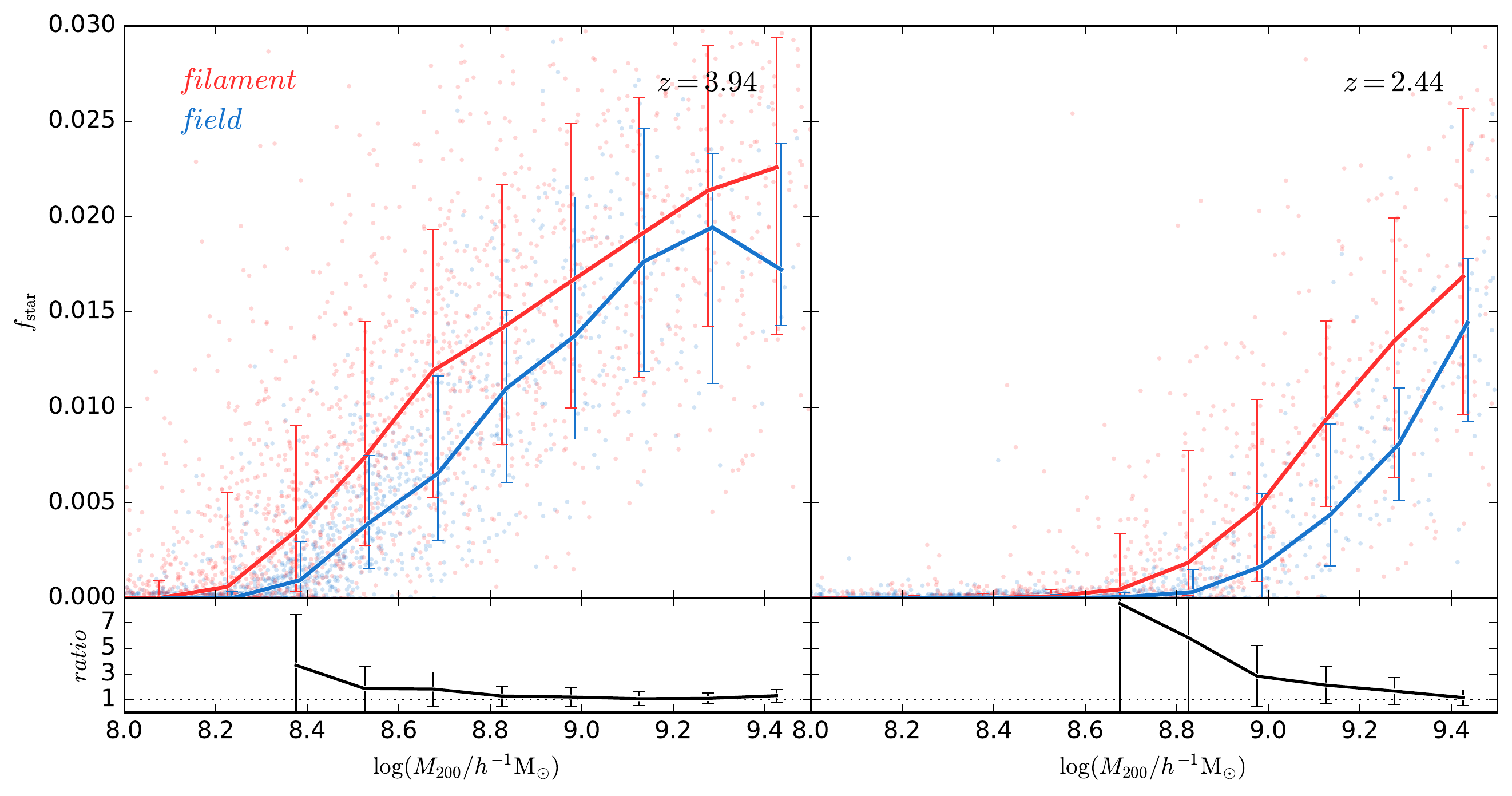}
	
	\caption{Similar to the Fig. \ref{fig:fig2}, but for the stellar fractions, $f_{\rm star}$. }
	
	\label{fig:fig3}
\end{figure*}

\section{Results}
\label{sec:results}
\subsection{Baryonic fractions and stellar fractions}

Using their simulations with a simple galaxy formation model, \citetalias{Liao&Gao2019} have shown that, compared to the dwarf galaxies in the field, the dwarfs in filaments with the same halo mass tend to have higher baryonic fractions and stellar fractions, suggesting that filaments play a significant role in assisting gas cooling and star formation. Since they studied with a simple hydrodynamic version of the Aquarius simulation \citep{Springel2008}, which does not include any feedback mechanism, here we re-examine these results with the Auriga simulations to see how these results vary in hydrodynamical simulations with more realistic galaxy formation models.

Following \citetalias{Liao&Gao2019}, we define the baryonic fraction of a galaxy as $f_{\rm bar} \equiv M_{\rm bar}/M_{200}$, where $M_{\rm bar}$ is the baryonic mass (including gas and stellar masses) within the virial radius $R_{200}$.
In Fig.~\ref{fig:fig2}, we display baryonic fractions for filament (red dots) and field galaxy (blue dots) of our galaxy sample at redshift $z=3.94$ (left panel) and $z=2.44$ (right panel) respectively, as the median values for filament and field galaxies are shown with red and blue lines, and the scatters (i.e. the 16th and 84th percentiles) in each mass bin are shown with the error bars. In the bottom panels, we plot the ratio between the median baryonic fraction of filament galaxies and that of field ones, and the error of the ratio is computed from error propagation, i.e. $\delta(x_{\rm fila}/x_{\rm field}) = \sqrt{(\delta x_{\rm fila} / x_{\rm field})^2 + (x_{\rm fila} \delta x_{\rm field} / x_{\rm field}^2)^2}$. Note that performing error propagation with asymmetric errors from the upper panels is not a trivial task \citep[see e.g.][]{Barlow2004,Audi2017,Possolo2019}. Here we first follow the `Method 2' in Appendix A of \citet{Audi2017} to estimate the equivalent symmetric errors (i.e. $\delta x_{\rm fila}$ and $\delta x_{\rm field}$) and median values (i.e. $x_{\rm fila}$ and $x_{\rm field}$) for both filament and field galaxies in each mass bin, then use the aforementioned error propagation formula to compute the error of the ratio.
Note that as there are very few field galaxies with $M_{200} \ga 10^{9.5}~h^{-1}{\rm M}_\odot$ in the Auriga simulations, here we only plot and focus on the galaxies with $M_{200} \leq 10^{9.5}~h^{-1}{\rm M}_\odot$ in both environments. 

During the cosmic reionization, as the ultraviolet (UV) background produced by quasars and stars heats the gas via the photoheating process, the gas becomes too hot to be confined by the gravitational potential wells of low-mass dark matter haloes, leading to a decrease in baryonic fractions in lower mass haloes. We observe here that, below $M_{200} \la 10^{9.5}~h^{-1}{\rm M}_\odot$, the baryonic fractions of galaxies in both filaments and field decrease with halo masses, in agreement with previous results \citep[e.g.][]{Okamoto2008}. We also note that the characteristic mass, defined as the galaxy virial mass at which the median baryonic fraction is half of the cosmic baryonic fraction ($\Omega_{\rm b}/\Omega_{\rm m}$), increases with time for both filament and field galaxies, i.e. the characteristic mass is $M_{\rm c} \sim 10^{8.5}~h^{-1}{\rm M}_\odot$ at $z \sim 4$ while it is $M_{\rm c} \sim 10^{9}~h^{-1}{\rm M}_\odot$ at $z \sim 2.5$. This redshift dependence of characteristic mass is mainly a result of reinonization and hierarchical halo assembly \citep[see][for a detailed semi-analytical model to explain this]{Okamoto2008}. We also expect that the supernova feedback, which is not included in \citet{Okamoto2008}, partially contributes to the evolution of characteristic mass as the feedback heats and enriches gas which consequently affects the heating and cooling processes and the gas equilibrium temperature. 

Overall, at fixed halo mass,  filament galaxies tend to have a higher baryonic fraction than those in the field, and this enhancement of the baryonic fraction in filament galaxies is larger with decreasing halo mass. At $M_{200} \sim 10^8~h^{-1}{\rm M}_\odot$, the median baryonic fraction for filament galaxies is $\sim 5$ times that for field ones. The effects are observed both at $z \approx 4$ and $z \approx 2.5$. 

Quantitatively, these results are very similar to those in \citetalias{Liao&Gao2019}. For example, in \citetalias{Liao&Gao2019}, at $z=4$, as the galaxy mass decreases from $\sim 10^9~h^{-1}{\rm M}_\odot$ to $10^{7.5}~h^{-1}{\rm M}_\odot$, the ratio between the median baryonic fractions of filament galaxies and field galaxies increases from $\sim 1.1$ to $\sim 5$; for galaxies with $M_{200} \sim 10^8~h^{-1}{\rm M}_\odot$, this ratio is $\sim 3$. This implies that adding supernovae feedback has a small effect on the 
{$f_{\rm bar}$-difference between filament and field high-$z$ dwarf galaxies}.

{We notice that there are a fraction of galaxies which have baryonic fractions higher than the cosmic mean value (i.e. $\Omega_{\rm b} / \Omega_{\rm m} \approx 0.16$), especially in the filament environment. One cause is the environmental effects, i.e. as the Auriga simulations focus on overdense zoom-in regions with stronger baryon inflows compared to the cosmic average, the baryonic fractions can be a little bit higher than the cosmic mean value, especially in the high-density filament environment. Another cause is that some haloes are splashback haloes, i.e. they used to be subhaloes in the history and some fraction of the outer dark matter mass was tidally stripped away and left with a higher baryonic fraction \citep[see similar discussions in][]{Okamoto2008}. To study the impacts from these splashback haloes on our results, we have excluded all galaxies which are within $3R_{200}$ of more massive galaxies, and found that the median $f_{\rm bar}$-$M_{200}$ relations for both our filament and field galaxies are marginally affected.}

In Fig.~\ref{fig:fig3}, we show stellar fractions of both filament and field galaxies as a function of their parent halo masses. Here the stellar fraction is  defined as $f_{\rm star} \equiv M_{\rm star}/M_{200}$ where $M_{\rm star}$ is the stellar mass inside the virial radius of a galaxy. 
{At both $z \approx 4$ and $z \approx 2.5$, compared to \citetalias{Liao&Gao2019}, for galaxies with the same virial masses, the stellar fractions in the Auriga simulations tend to be lower (e.g. for galaxies with $M_{200} \sim 10^9~h^{-1}{\rm M}_\odot$, the Auriga ones have a median stellar fraction of $\sim 1.5\%$ while the ones in \citetalias{Liao&Gao2019} have a median of $\sim 15\%$). This is a result of the supernovae feedback processes, i.e. as has been shown in many previous studies, supernovae-driven winds with variable wind velocities used in the Auriga simulations are an effective mechanism in simulations to suppress star formation in low-mass galaxies and to reproduce the observed stellar mass fractions and galaxy stellar mass function at the low-mass end  \citep[e.g.][]{Okamoto2010,Puchwein2013,Vogelsberger2013}. Interestingly, although the absolute $f_{\rm star}$ values of Auriga dwarf galaxies are lower than those of similar-mass galaxies in \citetalias{Liao&Gao2019}, we can still see that at both reshifts, the Auriga dwarf galaxies in filaments tend to have higher stellar masses than their counterparts in the field, and this environmental difference is larger for galaxies with lower halo masses, which is qualitatively similar to \citetalias{Liao&Gao2019}.}

In general, our results support the conclusions in \citetalias{Liao&Gao2019} that filaments play a role in assisting gas cooling and enhancing star formation. Similar trends have also been seen for galaxies with higher masses (e.g. virial masses $\ga 10^{10}~h^{-1}{\rm M}_\odot$) in hydrodynamical simulations in cosmological volumes \citep[e.g.][]{Metuki2015, Xu2020}. { For example, \citet{Xu2020} present the cosmic web dependence of baryonic/stellar fractions for galaxies with $M_{200} \geq 10^{10}~h^{-1}{\rm M}_\odot$ at $z = 0, 1, 3, 6$ using the EAGLE simulations \citep{Schaye2015}. They find that such cosmic web dependence of baryonic fractions increases with redshift up to $z=3$ at low masses, but the cosmic web dependence of stellar fractions in their simulations only exist at $z=0$ and $1$.} 
Overall, compared to the filament and void galaxies in these cosmological-box simulations, the differences between the filament and field galaxies in this study are more pronounced, possibly due to the fact that the filament environment has a stronger impact on galaxy formation in low mass haloes. 

In observations, some studies suggest that galaxies residing closer to filaments tend to have  higher stellar masses \citep[e.g.][]{Alpaslan2016, Poudel2017, Lee2020}, which is consistent with the trend in our simulations. However, we should stress that these studies are usually based on galaxies at much lower redshifts (e.g. $z \la 0.2$).

\begin{figure*}
    \centering
	\includegraphics[width=2.03\columnwidth]{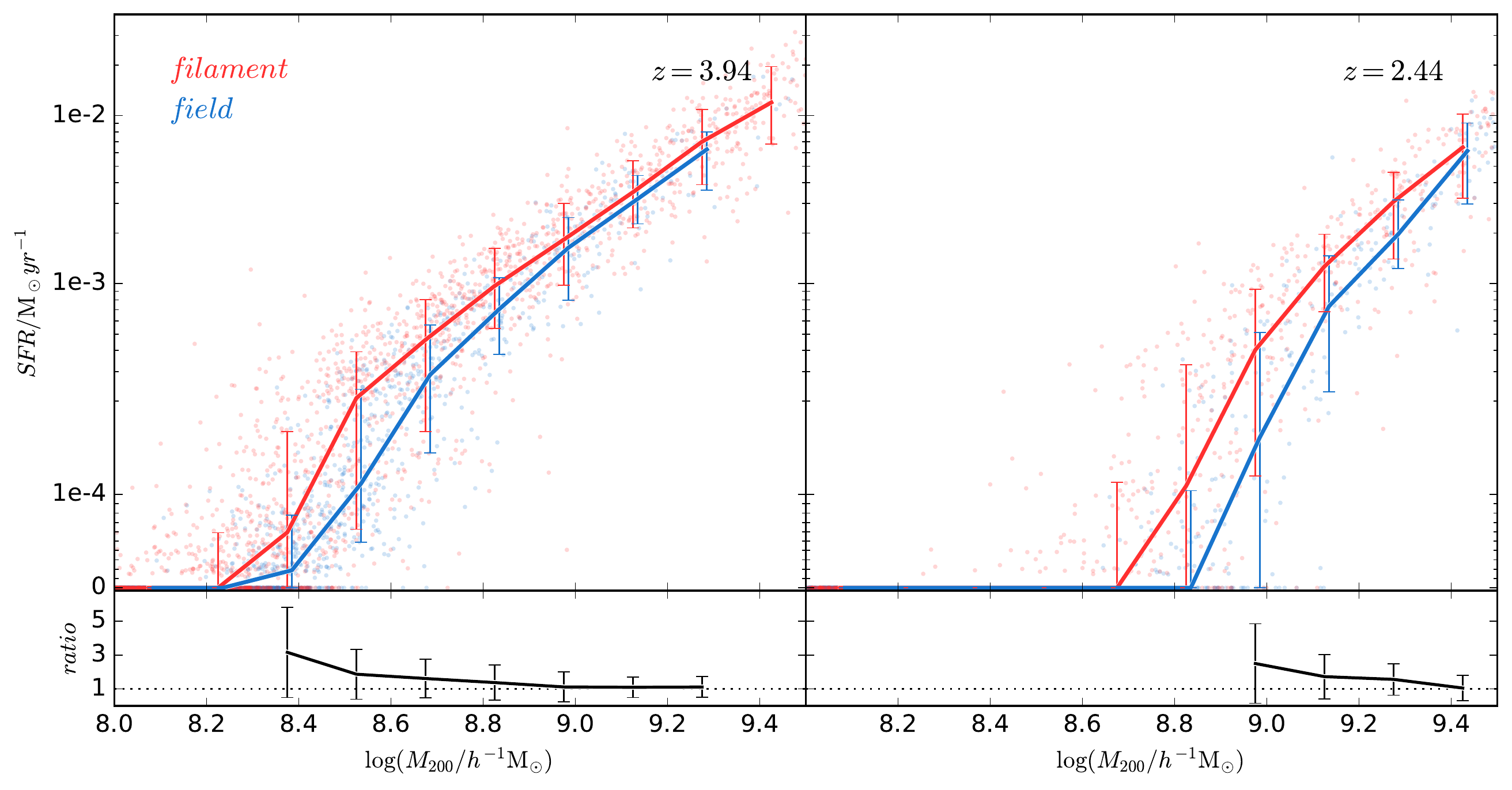}

    \caption{Similar to Fig. \ref{fig:fig2}, but for galaxy star formation rates (SFR). {Only galaxies with at least 20 gas cells from the parent sample are considered here. The medians and scatters are only shown for the mass bins containing at least 20 galaxies. Note that in the upper panels, in order to show all the data points, we use linear scale for $0 \leq {\rm SFR} \leq 10^{-4}$ ${\rm M}_\odot~{\rm yr}^{-1}$ and logarithmic scale for ${\rm SFR} > 10^{-4}$ ${\rm M}_\odot~{\rm yr}^{-1}$.}}
	
	\label{fig:fig4}
\end{figure*}

\begin{figure*}
    \centering
	\includegraphics[width=2.03\columnwidth]{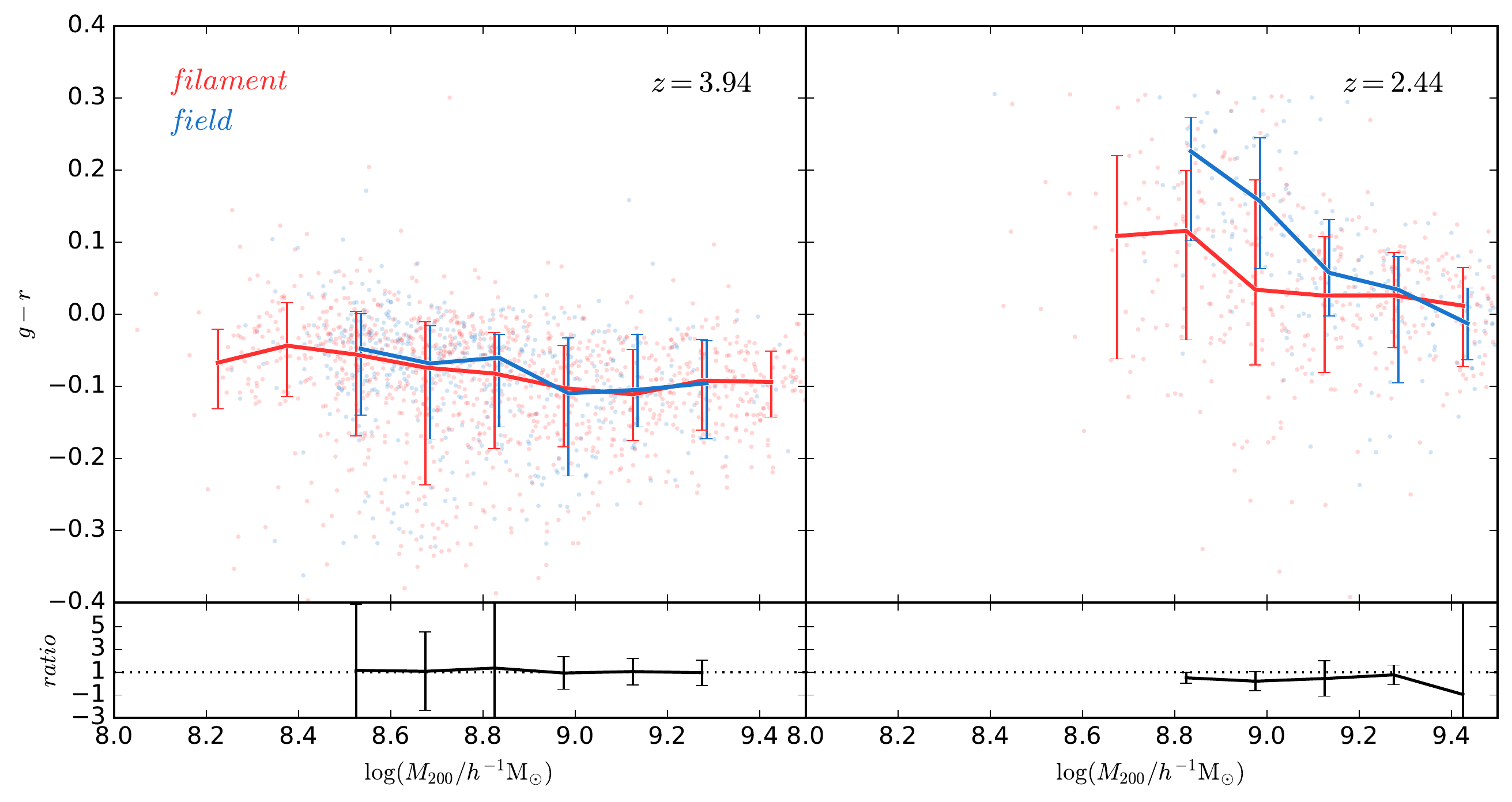}

    \caption{Similar to the Fig. \ref{fig:fig2}, but for galaxy colours $g-r$. {Only galaxies with at least 20 star particles from the parent sample are used here. The medians and error bars are only shown for the mass bins containing at least 20 galaxies.}}
	
	\label{fig:fig5}
\end{figure*}

\begin{figure*}
    \centering
	\includegraphics[width=2.03\columnwidth]{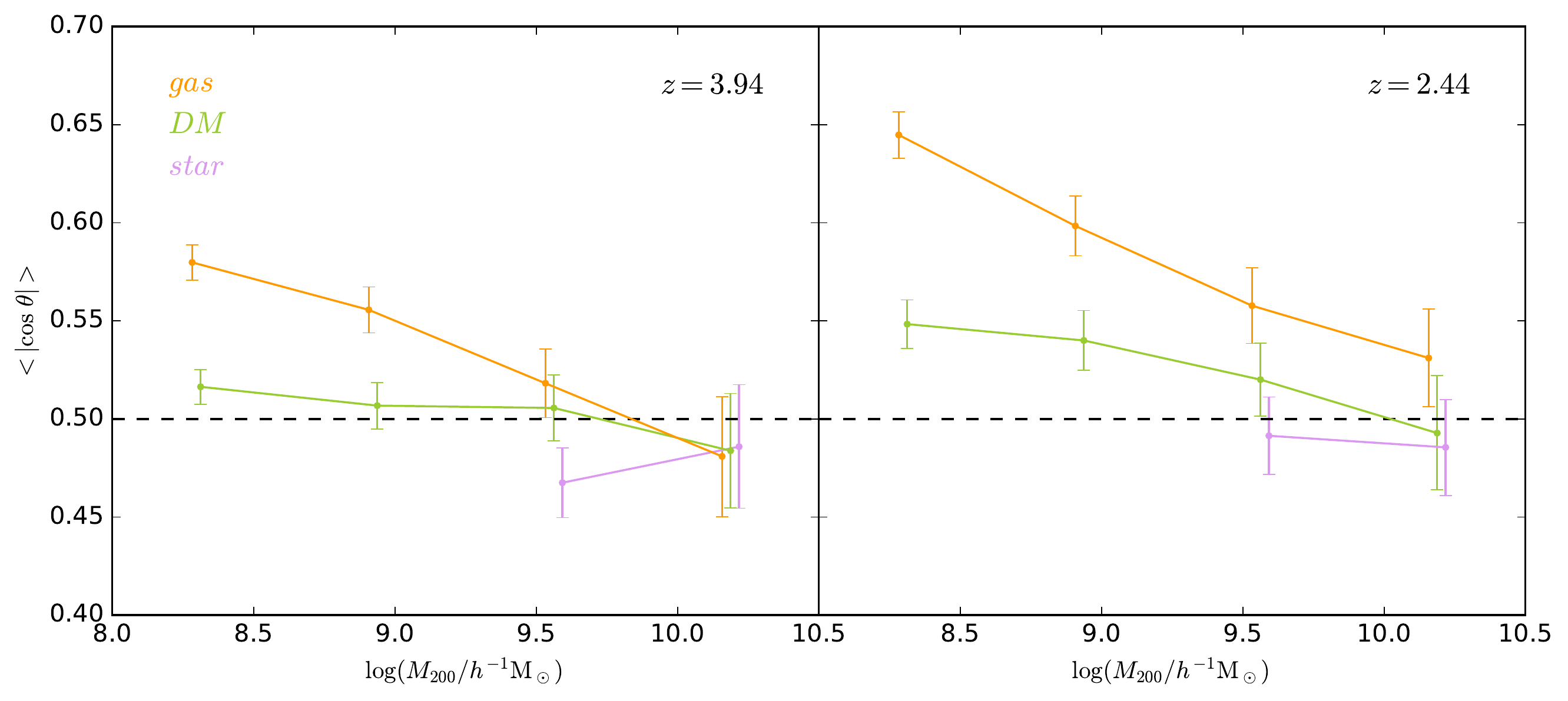}
	\caption{The relation between average $\lvert \cos \theta \rvert$ and galaxy virial mass, $M_{200}$, at $z=3.94$ (left) and $2.44$ (right). Orange, green and purple lines show the results of using {gas}, dark matter and star components to compute the spin directions, respectively.
	{The error bars show the 1$\sigma$ uncertainty estimated by bootstrap resampling; see the main text for details. To be clear, the data points of the dark matter and star components have been slightly shifted along the horizontal axis.}
	The dashed line shows the expected mean value for a uniform $\lvert \cos \theta \rvert$ distribution if no alignments at present.
	}
	
	\label{fig:fig6}
\end{figure*}

\subsection{Star formation rates and galaxy colours}

As shown above since filaments play a role in assisting gas cooling and star formation, we expect that dwarf galaxies residing in filaments may have higher star formation rates and thus bluer colours. To examine this, we plot the star formation rates ($g-r$ colours) of our galaxy sample as a function of halo mass for both filament and field haloes in Fig.~\ref{fig:fig4} (Fig.~\ref{fig:fig5}). {To avoid unreliable estimations due to poor resolution, we only consider galaxies with at least 20 gas cells (20 star particles) within $R_{200}$ from the parent sample when computing star formation rates ($g-r$ colours). Furthermore, we only show the median for those mass bins which contain at least 20 galaxies.} As expected, dwarf galaxies in filaments tend to have slightly higher star formation rates at both $z \approx 4$ and $z \approx 2.5$ compared to those with similar masses residing in the field. In particular, differences in star formation rates tends to be larger at lower redshifts ($z \sim 2.5$). Our results are consistent with \citet{Metuki2015} who also found that the lowest mass filament galaxies in their sample (i.e. $10^{10} - 10^{11}~h^{-1}{\rm M}_\odot$) at higher redshifts { (i.e. $z \ga 1$)} tend to have higher star formation rates than void galaxies. {However, apart from the two mass bins of $M_{200} \sim 10^{8.8}$ and $\sim 10^{9}~{\rm M}_\odot$ at $z \approx 2.5$ in which the filament galaxies tend to be slightly bluer, there is no clear difference in galaxy colours between the filament and field populations.} In \citet{Xu2020}, at $z=0$ void galaxies in their simulations tend to be slightly redder than filament galaxies with $M_{\rm star} \la 10^{10}~h^{-1}{\rm M}_\odot$, but such environmental dependence vanishes at $z \geq 1${, which is similar to our results here}. 
On the other hand, our results are inconsistent with some observational results on low redshift filament galaxies \citep[e.g. ][]{Alpaslan2016, Poudel2017, Odekon2018, Lee2020}, which found that galaxies residing closer to filaments tend to be redder and have lower star formation rates. This inconsistency might be due to that properties of the filament galaxies may be different at high and low redshift.

\subsection{Alignment between galaxy spins and filaments}

As shown in \citetalias{Liao&Gao2019} gas accretion in dwarf haloes in filaments is very anisotropic (i.e. the gas accreting along and perpendicular to the filament forms two distinct modes). It is interesting to study the alignment relation between filaments and their resident dwarfs. In previous studies, the alignment between galaxy orientations and filaments has been investigated with hydrodynamical simulations in cosmological volumes for massive galaxies at low redshift \citep[e.g.][]{Dubois2014,Codis2018,Wang2018,GaneshaiahVeena2019,Kraljic2020}. Here we extend these studies to lower mass dwarf galaxies and to higher redshift with the Auriga simulations.

The alignment between the orientation of a galaxy and the filament in which this galaxy resides can be described as
\begin{equation}
    \cos \theta = \frac{\bm{n}_{\rm spin} \cdot \bm{n}_{\rm fila}}{|\bm{n}_{\rm spin}| |\bm{n}_{\rm fila}|}.
\end{equation}
Here $\bm{n}_{\rm spin}$ is the spin direction of a galaxy, and $\bm{n}_{\rm fila}$ is the direction of the filament in which the galaxy resides. The former one is the same as the direction of the total angular momentum of particles inside $R_{200}$, and the latter is derived as follows. 

{ Following \citet{Hahn2007}, we use the Hessian matrix method to compute filament directions. The Hessian matrix of the smoothed density field at the galaxy centre ($\bm{x}_{\rm gal}$) is defined as}
\begin{equation}
    H_{ij} (\bm{x}_{\rm gal}) = \frac{\partial \rho(\bm{x}; R_{\rm s})}{\partial x_i \partial x_j} \left\vert_{\bm{x} = \bm{x}_{\rm gal}}\right. ,
\end{equation}
where $i,j = 1, 2, 3$ represent spatial dimensions, $\rho(\bm{x};R_{\rm s})$ is the smoothed filament density field (here the original filament density field is defined as the baryonic density field composed of the grid cells which are classified as filament cells; see panels (b) and (e) of Fig.~\ref{fig:fig1}), and the smoothing scale, $R_{\rm s} = 0.25 h^{-1}{\rm Mpc}$\footnote{We have tested changing the smoothing scale to $0.1$ and $0.5 h^{-1} \mathrm{Mpc}$, and confirmed that our results are not sensitive to the choice of smoothing scales.}. Here, the density field is smoothed by convolving with a Gaussian kernel in Fourier space according to \citet{Hahn2007}.
The filament direction $\bm{n}_{\rm fila}$ is then given by the eigenvector which is associated with the maximum eigenvalue of the Hessian matrix. See panels (c) and (f) of Fig.~\ref{fig:fig1} for examples of filament directions (cyan line segments) at the centres of filament galaxies.

In Fig.~\ref{fig:fig6}, we plot the alignments ($\left<|\cos \theta|\right>$) of different components (dark matter, {gas}, and stellar) of our filament galaxies as a function of their parent halo masses; median values {and the 1$\sigma$ errors} are shown. Following previous work \citep[e.g.][]{GaneshaiahVeena2019}, the 1$\sigma$ errors are estimated using the bootstrap resampling method. Specifically, in each mass bin, we generate $10^5$ realizations with the bootstrap method, compute the median $\left<|\cos \theta|\right>$ for each realization, and use the standard deviation of all these median $\left<|\cos \theta|\right>$ as an estimation of the 1$\sigma$ error. Note that in order to have a numerically robust estimation of the spin direction for the {gas} component, here we only consider filament galaxies which contain at least {50 gas cells} within the virial radius $R_{\mathrm{200}}$. We use a wider mass range $\left[10^{8.0}, 10^{10.5}\right]~h^{-1}{\rm M}_\odot$ here (as we do not need to consider the mass upper limit of the field sample), and divide it into 4 mass bins, {each of which contains $1108\ (542)$, $614\ (367)$, $278\ (239)$, $90\ (110)$ galaxies at $z=3.94\ (2.44)$}. Note that for the stellar component (purple line), we only use galaxies that contain more than 50 star particles [the sample numbers are $251\ (218)$ and $90\ (110)$ at two largest mass bins at $z=3.94$ ($z=2.44$)] and only show the results above $10^{9.25}h^{-1}\mathrm{M_{\odot}}$, which is the approximate value of the corresponding virial mass threshold. It is found that the averaged $|\cos \theta|$ of low mass dwarf galaxies (especially for the {gas} component and at $z \sim 2.5$) is larger than 0.5, suggesting that the direction of the spin tends to align with the direction of the filaments. This trend weakens (even reverses) at larger masses. 

This mass dependence of spin-filament alignment in baryonic and dark matter components is broadly in agreement with some previous hydrodynamical simulations \citep[see e.g.][]{Dubois2014, Codis2018, Wang2018, GaneshaiahVeena2019, Kraljic2020}, from which it was found that halo/stellar spins of low-mass galaxies tend to be parallel to their filaments while the halo (or gas) spins of high-mass galaxies tend to be perpendicular to filaments. 
There is also evidence that this transition mass decreases as redshifts increase \citep[e.g.][]{Wang2018}. But we should note that the spin of the stellar components of our galaxy samples (purple lines in the Fig.~\ref{fig:fig6}) does not show any clear parallel or perpendicular trend with respect to their filaments, while aforementioned studies \citep[e.g.][]{Wang2018, GaneshaiahVeena2019} suggest a slight perpendicular trend in larger galaxies ($M_{200} \ga 10^{11} h^{-1} \mathrm{M_{\odot}}$ or $M_{\mathrm{star}} \ga 10^{9} h^{-1} \mathrm{M_{\odot}}$). This is possibly due to the fact that we are looking at lower mass and higher redshift galaxies relative to hydrodynamical simulations, and also may to the lack of statistics (i.e. we only have two data points in a narrow mass range for stellar components at both redshifts in Fig.~\ref{fig:fig6}).

Note that the spin-filament alignment signal for (spiral) galaxies has also been reported in some observations at $z \la 0.2$ \citep[e.g.][]{Tempel&Libeskind2013, Tempel2013, BlueBird2020,Welker2020}, but these observational results are still quite controversial \citep[see e.g.][]{Zhang2015,Pahwa2016,Hirv2017,Krolewski2019}.

\section{Conclusion}
\label{sec:conclusion}

In this paper, we have made use of the Auriga project, a suite of 30 hydrodynamical zoom-in simulations of Milky Way-like galaxies with a more realistic galaxy formation model, to study the impact of high redshift dense filaments on their resident dwarf galaxies.  

We find that compared to their counterparts in the field, dwarf galaxies residing in filaments tend to have higher baryonic and stellar fractions, and tend to have slightly higher star formation rates at both $z\sim 4$ and $z\sim 2.5$, in line with \citetalias{Liao&Gao2019} who carried out a similar study with a hydrodynamical simulation with a very simple galaxy formation model without taking into account of feedback (though the baryonic and stellar fractions in each environment are generally suppressed by feedbacks comparing to \citetalias{Liao&Gao2019}).
We also compare the galaxy colours of the filament and field populations, and they do not show clear difference. Overall, we conclude that dense and massive filaments can enhance the star formation of their resident dwarf galaxies. We also show that the gas in low-mass filament galaxies at high redshifts tend to have their spins aligned with the filament in which they reside, echoing the spin-filament signal observed for galaxies with higher masses and lower redshift seen in other hydrodynamical simulations.

\section*{Acknowledgements}
We thank the anonymous referee for a very constructive and useful report which helped to significantly improve our manuscript. We would like to thank Carlos S. Frenk for helpful discussions and advice, which significantly improved our manuscript. We acknowledge support from NSFC grants (Nos 11988101, 11133003, 11425312, 11903043, 12033008, 11622325), National Key Program for Science and Technology Research Development (2018YFA0404503, 2017YFB0203300), and K. C. Wong Foundation. SL acknowledges the support by the European Research Council via ERC Consolidator Grant KETJU (No. 818930). LG acknowledges the hospitality of the Institute for Computational Cosmology, Durham University. RG acknowledges financial support from the Spanish Ministry of Science and Innovation (MICINN) through the Spanish State Research Agency, under the Severo Ochoa Program 2020-2023 (CEX2019-000920-S). 

\section*{Data availability}
The data underlying this article will be shared on reasonable request to the corresponding author.

\bibliographystyle{mnras}
\bibliography{main} 

\begin{thebibliography}{}
\makeatletter
\relax
\def\mn@urlcharsother{\let\do\@makeother \do\$\do\&\do\#\do\^\do\_\do\%\do\~}
\def\mn@doi{\begingroup\mn@urlcharsother \@ifnextchar [ {\mn@doi@}
  {\mn@doi@[]}}
\def\mn@doi@[#1]#2{\def\@tempa{#1}\ifx\@tempa\@empty \href
  {http://dx.doi.org/#2} {doi:#2}\else \href {http://dx.doi.org/#2} {#1}\fi
  \endgroup}
\def\mn@eprint#1#2{\mn@eprint@#1:#2::\@nil}
\def\mn@eprint@arXiv#1{\href {http://arxiv.org/abs/#1} {{\tt arXiv:#1}}}
\def\mn@eprint@dblp#1{\href {http://dblp.uni-trier.de/rec/bibtex/#1.xml}
  {dblp:#1}}
\def\mn@eprint@#1:#2:#3:#4\@nil{\def\@tempa {#1}\def\@tempb {#2}\def\@tempc
  {#3}\ifx \@tempc \@empty \let \@tempc \@tempb \let \@tempb \@tempa \fi \ifx
  \@tempb \@empty \def\@tempb {arXiv}\fi \@ifundefined
  {mn@eprint@\@tempb}{\@tempb:\@tempc}{\expandafter \expandafter \csname
  mn@eprint@\@tempb\endcsname \expandafter{\@tempc}}}

\bibitem[\protect\citeauthoryear{{Alpaslan} et~al.,}{{Alpaslan}
  et~al.}{2016}]{Alpaslan2016}
{Alpaslan} M.,  et~al., 2016, \mn@doi [\mnras] {10.1093/mnras/stw134}, \href
  {https://ui.adsabs.harvard.edu/abs/2016MNRAS.457.2287A} {457, 2287}

\bibitem[\protect\citeauthoryear{{Audi}, {Kondev}, {Wang}, {Huang}  \&
  {Naimi}}{{Audi} et~al.}{2017}]{Audi2017}
{Audi} G.,  {Kondev} F.~G.,  {Wang} M.,  {Huang} W.~J.,   {Naimi} S.,  2017,
  \mn@doi [Chinese Physics C] {10.1088/1674-1137/41/3/030001}, \href
  {https://ui.adsabs.harvard.edu/abs/2017ChPhC..41c0001A} {41, 030001}

\bibitem[\protect\citeauthoryear{{Barlow}}{{Barlow}}{2004}]{Barlow2004}
{Barlow} R.,  2004, arXiv e-prints, \href
  {https://ui.adsabs.harvard.edu/abs/2004physics...6120B} {p. physics/0406120}

\bibitem[\protect\citeauthoryear{{Blue Bird} et~al.,}{{Blue Bird}
  et~al.}{2020}]{BlueBird2020}
{Blue Bird} J.,  et~al., 2020, \mn@doi [\mnras] {10.1093/mnras/stz3357}, \href
  {https://ui.adsabs.harvard.edu/abs/2020MNRAS.492..153B} {492, 153}

\bibitem[\protect\citeauthoryear{{Brooks}, {Governato}, {Quinn}, {Brook}  \&
  {Wadsley}}{{Brooks} et~al.}{2009}]{Brooks2009}
{Brooks} A.~M.,  {Governato} F.,  {Quinn} T.,  {Brook} C.~B.,   {Wadsley} J.,
  2009, \mn@doi [\apj] {10.1088/0004-637X/694/1/396}, \href
  {https://ui.adsabs.harvard.edu/abs/2009ApJ...694..396B} {694, 396}

\bibitem[\protect\citeauthoryear{{Codis}, {Jindal}, {Chisari}, {Vibert},
  {Dubois}, {Pichon}  \& {Devriendt}}{{Codis} et~al.}{2018}]{Codis2018}
{Codis} S.,  {Jindal} A.,  {Chisari} N.~E.,  {Vibert} D.,  {Dubois} Y.,
  {Pichon} C.,   {Devriendt} J.,  2018, \mn@doi [\mnras]
  {10.1093/mnras/sty2567}, \href
  {https://ui.adsabs.harvard.edu/abs/2018MNRAS.481.4753C} {481, 4753}

\bibitem[\protect\citeauthoryear{{Crain}, {Eke}, {Frenk}, {Jenkins},
  {McCarthy}, {Navarro}  \& {Pearce}}{{Crain} et~al.}{2007}]{Crain2007}
{Crain} R.~A.,  {Eke} V.~R.,  {Frenk} C.~S.,  {Jenkins} A.,  {McCarthy} I.~G.,
  {Navarro} J.~F.,   {Pearce} F.~R.,  2007, \mn@doi [\mnras]
  {10.1111/j.1365-2966.2007.11598.x}, \href
  {https://ui.adsabs.harvard.edu/abs/2007MNRAS.377...41C} {377, 41}

\bibitem[\protect\citeauthoryear{{Crone Odekon}, {Hallenbeck}, {Haynes},
  {Koopmann}, {Phi}  \& {Wolfe}}{{Crone Odekon} et~al.}{2018}]{Odekon2018}
{Crone Odekon} M.,  {Hallenbeck} G.,  {Haynes} M.~P.,  {Koopmann} R.~A.,  {Phi}
  A.,   {Wolfe} P.-F.,  2018, \mn@doi [\apj] {10.3847/1538-4357/aaa1e8}, \href
  {https://ui.adsabs.harvard.edu/abs/2018ApJ...852..142C} {852, 142}

\bibitem[\protect\citeauthoryear{{Davis}, {Efstathiou}, {Frenk}  \&
  {White}}{{Davis} et~al.}{1985}]{Davis1985}
{Davis} M.,  {Efstathiou} G.,  {Frenk} C.~S.,   {White} S.~D.~M.,  1985,
  \mn@doi [\apj] {10.1086/163168}, \href
  {https://ui.adsabs.harvard.edu/abs/1985ApJ...292..371D} {292, 371}

\bibitem[\protect\citeauthoryear{{Dekel} \& {Birnboim}}{{Dekel} \&
  {Birnboim}}{2006}]{Dekel2006}
{Dekel} A.,  {Birnboim} Y.,  2006, \mn@doi [\mnras]
  {10.1111/j.1365-2966.2006.10145.x}, \href
  {https://ui.adsabs.harvard.edu/abs/2006MNRAS.368....2D} {368, 2}

\bibitem[\protect\citeauthoryear{{Dekel} et~al.,}{{Dekel}
  et~al.}{2009}]{Dekel2009}
{Dekel} A.,  et~al., 2009, \mn@doi [\nat] {10.1038/nature07648}, \href
  {https://ui.adsabs.harvard.edu/abs/2009Natur.457..451D} {457, 451}

\bibitem[\protect\citeauthoryear{{Dubois} et~al.,}{{Dubois}
  et~al.}{2014}]{Dubois2014}
{Dubois} Y.,  et~al., 2014, \mn@doi [\mnras] {10.1093/mnras/stu1227}, \href
  {https://ui.adsabs.harvard.edu/abs/2014MNRAS.444.1453D} {444, 1453}

\bibitem[\protect\citeauthoryear{{Ganeshaiah Veena}, {Cautun}, {Tempel}, {van
  de Weygaert}  \& {Frenk}}{{Ganeshaiah Veena}
  et~al.}{2019}]{GaneshaiahVeena2019}
{Ganeshaiah Veena} P.,  {Cautun} M.,  {Tempel} E.,  {van de Weygaert} R.,
  {Frenk} C.~S.,  2019, \mn@doi [\mnras] {10.1093/mnras/stz1343}, \href
  {https://ui.adsabs.harvard.edu/abs/2019MNRAS.487.1607G} {487, 1607}

\bibitem[\protect\citeauthoryear{{Gao} \& {Theuns}}{{Gao} \&
  {Theuns}}{2007}]{Gao2007}
{Gao} L.,  {Theuns} T.,  2007, \mn@doi [Science] {10.1126/science.1146676},
  \href {https://ui.adsabs.harvard.edu/abs/2007Sci...317.1527G} {317, 1527}

\bibitem[\protect\citeauthoryear{{Gao}, {Theuns}  \& {Springel}}{{Gao}
  et~al.}{2015}]{Gao2015}
{Gao} L.,  {Theuns} T.,   {Springel} V.,  2015, \mn@doi [\mnras]
  {10.1093/mnras/stv643}, \href
  {https://ui.adsabs.harvard.edu/abs/2015MNRAS.450...45G} {450, 45}

\bibitem[\protect\citeauthoryear{{Grand} et~al.,}{{Grand}
  et~al.}{2017}]{Grand2017}
{Grand} R.~J.~J.,  et~al., 2017, \mn@doi [\mnras] {10.1093/mnras/stx071}, \href
  {https://ui.adsabs.harvard.edu/abs/2017MNRAS.467..179G} {467, 179}

\bibitem[\protect\citeauthoryear{{Hahn}, {Porciani}, {Carollo}  \&
  {Dekel}}{{Hahn} et~al.}{2007}]{Hahn2007}
{Hahn} O.,  {Porciani} C.,  {Carollo} C.~M.,   {Dekel} A.,  2007, \mn@doi
  [\mnras] {10.1111/j.1365-2966.2006.11318.x}, \href
  {https://ui.adsabs.harvard.edu/abs/2007MNRAS.375..489H} {375, 489}

\bibitem[\protect\citeauthoryear{{Hirv}, {Pelt}, {Saar}, {Tago}, {Tamm},
  {Tempel}  \& {Einasto}}{{Hirv} et~al.}{2017}]{Hirv2017}
{Hirv} A.,  {Pelt} J.,  {Saar} E.,  {Tago} E.,  {Tamm} A.,  {Tempel} E.,
  {Einasto} M.,  2017, \mn@doi [\aap] {10.1051/0004-6361/201629248}, \href
  {https://ui.adsabs.harvard.edu/abs/2017A&A...599A..31H} {599, A31}

\bibitem[\protect\citeauthoryear{{Hoshen} \& {Kopelman}}{{Hoshen} \&
  {Kopelman}}{1976}]{Hoshen&Kopelman1976}
{Hoshen} J.,  {Kopelman} R.,  1976, \mn@doi [\prb] {10.1103/PhysRevB.14.3438},
  \href {https://ui.adsabs.harvard.edu/abs/1976PhRvB..14.3438H} {14, 3438}

\bibitem[\protect\citeauthoryear{{Kere{\v{s}}}, {Katz}, {Weinberg}  \&
  {Dav{\'e}}}{{Kere{\v{s}}} et~al.}{2005}]{Keres2005}
{Kere{\v{s}}} D.,  {Katz} N.,  {Weinberg} D.~H.,   {Dav{\'e}} R.,  2005,
  \mn@doi [\mnras] {10.1111/j.1365-2966.2005.09451.x}, \href
  {https://ui.adsabs.harvard.edu/abs/2005MNRAS.363....2K} {363, 2}

\bibitem[\protect\citeauthoryear{{Kere{\v{s}}}, {Katz}, {Fardal}, {Dav{\'e}}
  \& {Weinberg}}{{Kere{\v{s}}} et~al.}{2009}]{Keres2009}
{Kere{\v{s}}} D.,  {Katz} N.,  {Fardal} M.,  {Dav{\'e}} R.,   {Weinberg} D.~H.,
   2009, \mn@doi [\mnras] {10.1111/j.1365-2966.2009.14541.x}, \href
  {https://ui.adsabs.harvard.edu/abs/2009MNRAS.395..160K} {395, 160}

\bibitem[\protect\citeauthoryear{{Kraljic}, {Dav{\'e}}  \& {Pichon}}{{Kraljic}
  et~al.}{2020}]{Kraljic2020}
{Kraljic} K.,  {Dav{\'e}} R.,   {Pichon} C.,  2020, \mn@doi [\mnras]
  {10.1093/mnras/staa250}, \href
  {https://ui.adsabs.harvard.edu/abs/2020MNRAS.493..362K} {493, 362}

\bibitem[\protect\citeauthoryear{{Krolewski}, {Ho}, {Chen}, {Chan}, {Tenneti},
  {Bizyaev}  \& {Kraljic}}{{Krolewski} et~al.}{2019}]{Krolewski2019}
{Krolewski} A.,  {Ho} S.,  {Chen} Y.-C.,  {Chan} P.~F.,  {Tenneti} A.,
  {Bizyaev} D.,   {Kraljic} K.,  2019, \mn@doi [\apj]
  {10.3847/1538-4357/ab1010}, \href
  {https://ui.adsabs.harvard.edu/abs/2019ApJ...876...52K} {876, 52}

\bibitem[\protect\citeauthoryear{{Lee}, {Kim}, {Rey}  \& {Chung}}{{Lee}
  et~al.}{2021}]{Lee2020}
{Lee} Y.,  {Kim} S.,  {Rey} S.-C.,   {Chung} J.,  2021, \mn@doi [\apj]
  {10.3847/1538-4357/abcaa0}, \href
  {https://ui.adsabs.harvard.edu/abs/2021ApJ...906...68L} {906, 68}

\bibitem[\protect\citeauthoryear{{Liao} \& {Gao}}{{Liao} \&
  {Gao}}{2019}]{Liao&Gao2019}
{Liao} S.,  {Gao} L.,  2019, \mn@doi [\mnras] {10.1093/mnras/stz441}, \href
  {https://ui.adsabs.harvard.edu/abs/2019MNRAS.485..464L} {485, 464}

\bibitem[\protect\citeauthoryear{{Metuki}, {Libeskind}, {Hoffman}, {Crain}  \&
  {Theuns}}{{Metuki} et~al.}{2015}]{Metuki2015}
{Metuki} O.,  {Libeskind} N.~I.,  {Hoffman} Y.,  {Crain} R.~A.,   {Theuns} T.,
  2015, \mn@doi [\mnras] {10.1093/mnras/stu2166}, \href
  {https://ui.adsabs.harvard.edu/abs/2015MNRAS.446.1458M} {446, 1458}

\bibitem[\protect\citeauthoryear{{Ocvirk}, {Pichon}  \& {Teyssier}}{{Ocvirk}
  et~al.}{2008}]{Ocvirk2008}
{Ocvirk} P.,  {Pichon} C.,   {Teyssier} R.,  2008, \mn@doi [\mnras]
  {10.1111/j.1365-2966.2008.13763.x}, \href
  {https://ui.adsabs.harvard.edu/abs/2008MNRAS.390.1326O} {390, 1326}

\bibitem[\protect\citeauthoryear{{Okamoto}, {Gao}  \& {Theuns}}{{Okamoto}
  et~al.}{2008}]{Okamoto2008}
{Okamoto} T.,  {Gao} L.,   {Theuns} T.,  2008, \mn@doi [\mnras]
  {10.1111/j.1365-2966.2008.13830.x}, \href
  {https://ui.adsabs.harvard.edu/abs/2008MNRAS.390..920O} {390, 920}

\bibitem[\protect\citeauthoryear{{Okamoto}, {Frenk}, {Jenkins}  \&
  {Theuns}}{{Okamoto} et~al.}{2010}]{Okamoto2010}
{Okamoto} T.,  {Frenk} C.~S.,  {Jenkins} A.,   {Theuns} T.,  2010, \mn@doi
  [\mnras] {10.1111/j.1365-2966.2010.16690.x}, \href
  {https://ui.adsabs.harvard.edu/abs/2010MNRAS.406..208O} {406, 208}

\bibitem[\protect\citeauthoryear{{Pahwa} et~al.,}{{Pahwa}
  et~al.}{2016}]{Pahwa2016}
{Pahwa} I.,  et~al., 2016, \mn@doi [\mnras] {10.1093/mnras/stv2930}, \href
  {https://ui.adsabs.harvard.edu/abs/2016MNRAS.457..695P} {457, 695}

\bibitem[\protect\citeauthoryear{{Planck Collaboration} et~al.,}{{Planck
  Collaboration} et~al.}{2014}]{Planck2014p16}
{Planck Collaboration} et~al., 2014, \mn@doi [\aap]
  {10.1051/0004-6361/201321591}, \href
  {https://ui.adsabs.harvard.edu/abs/2014A%26A...571A..16P} {571, A16}

\bibitem[\protect\citeauthoryear{{Possolo}, {Merkatas}  \& {Bodnar}}{{Possolo}
  et~al.}{2019}]{Possolo2019}
{Possolo} A.,  {Merkatas} C.,   {Bodnar} O.,  2019, \mn@doi [Metrologia]
  {10.1088/1681-7575/ab2a8d}, \href
  {https://ui.adsabs.harvard.edu/abs/2019Metro..56d5009P} {56, 045009}

\bibitem[\protect\citeauthoryear{{Poudel}, {Hein{\"a}m{\"a}ki}, {Tempel},
  {Einasto}, {Lietzen}  \& {Nurmi}}{{Poudel} et~al.}{2017}]{Poudel2017}
{Poudel} A.,  {Hein{\"a}m{\"a}ki} P.,  {Tempel} E.,  {Einasto} M.,  {Lietzen}
  H.,   {Nurmi} P.,  2017, \mn@doi [\aap] {10.1051/0004-6361/201629639}, \href
  {https://ui.adsabs.harvard.edu/abs/2017A&A...597A..86P} {597, A86}

\bibitem[\protect\citeauthoryear{{Puchwein} \& {Springel}}{{Puchwein} \&
  {Springel}}{2013}]{Puchwein2013}
{Puchwein} E.,  {Springel} V.,  2013, \mn@doi [\mnras] {10.1093/mnras/sts243},
  \href {https://ui.adsabs.harvard.edu/abs/2013MNRAS.428.2966P} {428, 2966}

\bibitem[\protect\citeauthoryear{{Schaye} et~al.,}{{Schaye}
  et~al.}{2015}]{Schaye2015}
{Schaye} J.,  et~al., 2015, \mn@doi [\mnras] {10.1093/mnras/stu2058}, \href
  {https://ui.adsabs.harvard.edu/abs/2015MNRAS.446..521S} {446, 521}

\bibitem[\protect\citeauthoryear{{Springel}}{{Springel}}{2010}]{Springel2010}
{Springel} V.,  2010, \mn@doi [\mnras] {10.1111/j.1365-2966.2009.15715.x},
  \href {https://ui.adsabs.harvard.edu/abs/2010MNRAS.401..791S} {401, 791}

\bibitem[\protect\citeauthoryear{{Springel}, {White}, {Tormen}  \&
  {Kauffmann}}{{Springel} et~al.}{2001}]{Springel2001}
{Springel} V.,  {White} S.~D.~M.,  {Tormen} G.,   {Kauffmann} G.,  2001,
  \mn@doi [\mnras] {10.1046/j.1365-8711.2001.04912.x}, \href
  {https://ui.adsabs.harvard.edu/abs/2001MNRAS.328..726S} {328, 726}

\bibitem[\protect\citeauthoryear{{Springel} et~al.,}{{Springel}
  et~al.}{2008}]{Springel2008}
{Springel} V.,  et~al., 2008, \mn@doi [\mnras]
  {10.1111/j.1365-2966.2008.14066.x}, \href
  {https://ui.adsabs.harvard.edu/abs/2008MNRAS.391.1685S} {391, 1685}

\bibitem[\protect\citeauthoryear{{Tempel} \& {Libeskind}}{{Tempel} \&
  {Libeskind}}{2013}]{Tempel&Libeskind2013}
{Tempel} E.,  {Libeskind} N.~I.,  2013, \mn@doi [\apjl]
  {10.1088/2041-8205/775/2/L42}, \href
  {https://ui.adsabs.harvard.edu/abs/2013ApJ...775L..42T} {775, L42}

\bibitem[\protect\citeauthoryear{{Tempel}, {Stoica}  \& {Saar}}{{Tempel}
  et~al.}{2013}]{Tempel2013}
{Tempel} E.,  {Stoica} R.~S.,   {Saar} E.,  2013, \mn@doi [\mnras]
  {10.1093/mnras/sts162}, \href
  {https://ui.adsabs.harvard.edu/abs/2013MNRAS.428.1827T} {428, 1827}

\bibitem[\protect\citeauthoryear{{Vogelsberger}, {Genel}, {Sijacki}, {Torrey},
  {Springel}  \& {Hernquist}}{{Vogelsberger} et~al.}{2013}]{Vogelsberger2013}
{Vogelsberger} M.,  {Genel} S.,  {Sijacki} D.,  {Torrey} P.,  {Springel} V.,
  {Hernquist} L.,  2013, \mn@doi [\mnras] {10.1093/mnras/stt1789}, \href
  {https://ui.adsabs.harvard.edu/abs/2013MNRAS.436.3031V} {436, 3031}

\bibitem[\protect\citeauthoryear{{Wang}, {Guo}, {Kang}  \& {Libeskind}}{{Wang}
  et~al.}{2018}]{Wang2018}
{Wang} P.,  {Guo} Q.,  {Kang} X.,   {Libeskind} N.~I.,  2018, \mn@doi [\apj]
  {10.3847/1538-4357/aae20f}, \href
  {https://ui.adsabs.harvard.edu/abs/2018ApJ...866..138W} {866, 138}

\bibitem[\protect\citeauthoryear{{Welker} et~al.,}{{Welker}
  et~al.}{2020}]{Welker2020}
{Welker} C.,  et~al., 2020, \mn@doi [\mnras] {10.1093/mnras/stz2860}, \href
  {https://ui.adsabs.harvard.edu/abs/2020MNRAS.491.2864W} {491, 2864}

\bibitem[\protect\citeauthoryear{{White} \& {Frenk}}{{White} \&
  {Frenk}}{1991}]{White&Frenk1991}
{White} S. D.~M.,  {Frenk} C.~S.,  1991, \mn@doi [\apj] {10.1086/170483}, \href
  {https://ui.adsabs.harvard.edu/abs/1991ApJ...379...52W} {379, 52}

\bibitem[\protect\citeauthoryear{{White} \& {Rees}}{{White} \&
  {Rees}}{1978}]{White&Rees1978}
{White} S.~D.~M.,  {Rees} M.~J.,  1978, \mn@doi [\mnras]
  {10.1093/mnras/183.3.341}, \href
  {https://ui.adsabs.harvard.edu/abs/1978MNRAS.183..341W} {183, 341}

\bibitem[\protect\citeauthoryear{{Xu} et~al.,}{{Xu} et~al.}{2020}]{Xu2020}
{Xu} W.,  et~al., 2020, \mn@doi [\mnras] {10.1093/mnras/staa2497}, \href
  {https://ui.adsabs.harvard.edu/abs/2020MNRAS.498.1839X} {498, 1839}

\bibitem[\protect\citeauthoryear{{Zhang}, {Yang}, {Wang}, {Wang}, {Luo}, {Mo}
  \& {van den Bosch}}{{Zhang} et~al.}{2015}]{Zhang2015}
{Zhang} Y.,  {Yang} X.,  {Wang} H.,  {Wang} L.,  {Luo} W.,  {Mo} H.~J.,   {van
  den Bosch} F.~C.,  2015, \mn@doi [\apj] {10.1088/0004-637X/798/1/17}, \href
  {https://ui.adsabs.harvard.edu/abs/2015ApJ...798...17Z} {798, 17}

\makeatother
\end{thebibliography}

\appendix

\section{Resolution studies}\label{ap:resolution}

We mainly presented the results for 30 Level 4 (L4) Auriga simulations in the main text. The Auriga project has performed six Level 3 (L3) simulation (i.e. Au-6, Au-16, Au-21, Au-23, Au-24, and Au-27) which have higher resolution (i.e. $m_{\rm DM} = 4 \times 10^4~ {\rm M}_\odot$ and $m_{\rm b} = 6 \times 10^3~{\rm M}_\odot$). To see whether our results in the main text depend on numerical resolution, we have also analysed these six L3 simulations and compared their results with those from the corresponding six L4 simulations.

\begin{figure}
    \centering
    \includegraphics[width=1.0\columnwidth]{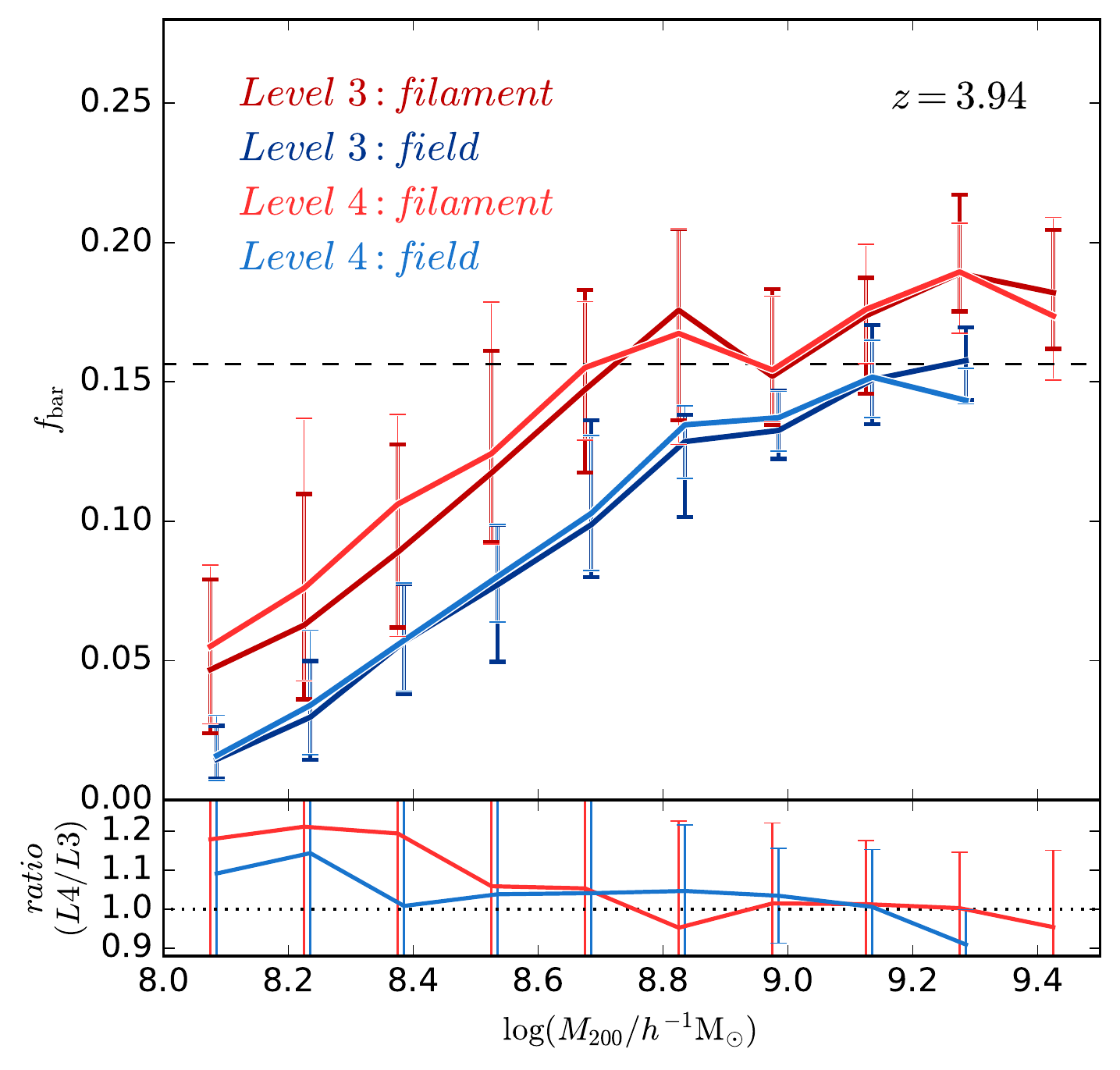}
    \includegraphics[width=1.0\columnwidth]{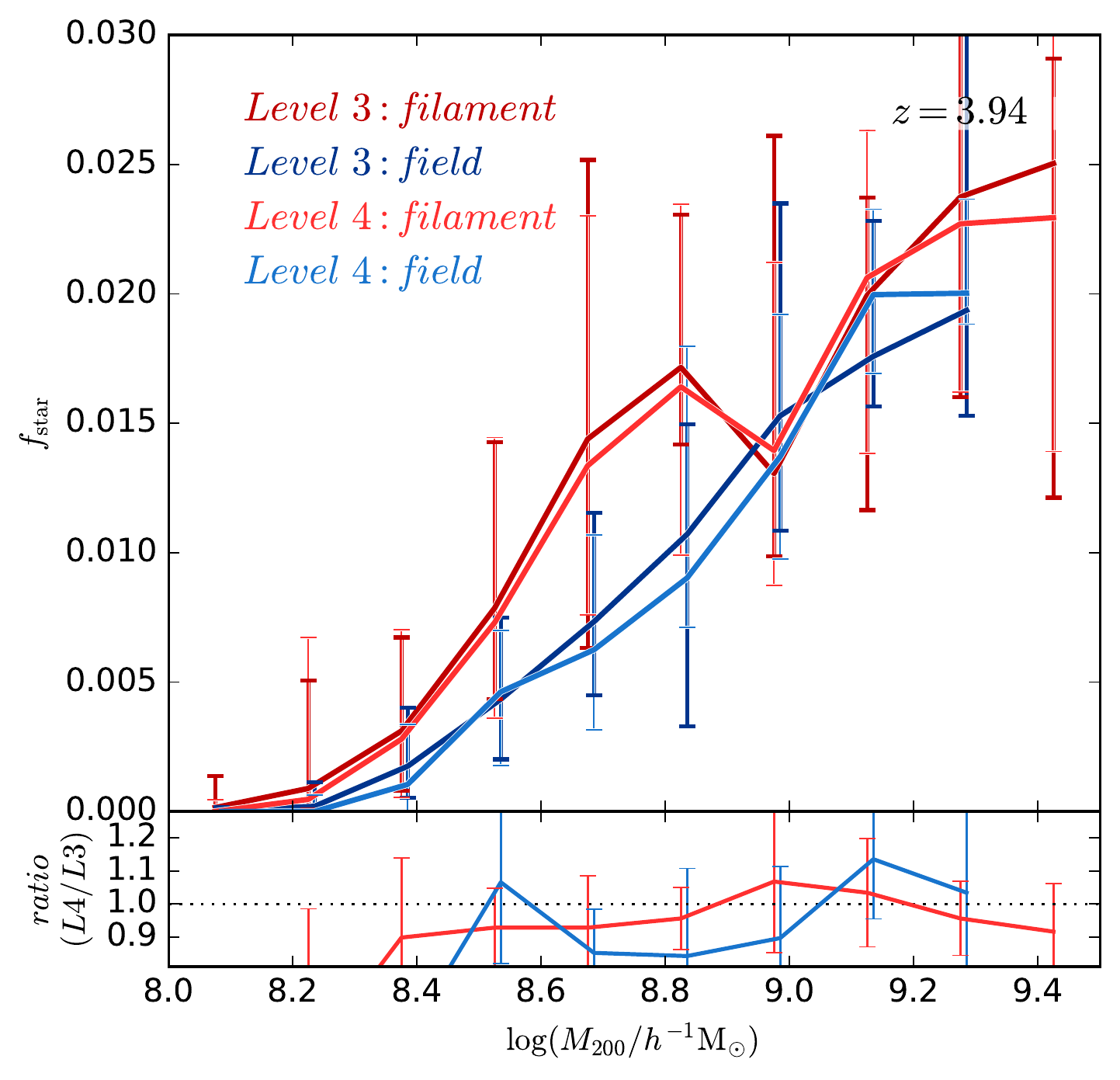}
    \caption{{Top:} Resolution convergence of $f_{\rm bar}$-$M_{200}$ relations. In the upper subpanel, the solid lines show the median relations at $z = 3.94$ for filament (red) and field (blue) galaxies identified from L4 (light color) and L3 (dark color) simulations, and the error bars show the 16th and 84th percentiles in each mass bin. The lower subpanel shows the ratio between the L4 and L3 median relations, and the errors of ratios are computed in the same way as those in Fig.~\ref{fig:fig2}. {Bottom:} Similar to the top panel, but for the resolution convergence of $f_{\rm star}$-$M_{200}$ relations.}
    \label{fig:figA1}
\end{figure}

In the top panel of Fig.~\ref{fig:figA1}, we compare the median baryonic fraction - virial mass relations at $z \sim 4$ for both filament and field galaxies from the L4 and L3 simulations. We can see that the relations from the L4 simulations agree fairly well with those from the L3 simulations, i.e. the differences are $\la 20\%$ in all mass bins. Especially, the difference in $f_{\rm bar}$ between filament and field galaxies in each mass bin is quite robust in simulations with different resolution, suggesting that our main conclusions are not sensitive to numerical resolution. Similar convergence results can be found for other median relations at both $z \sim 4$ and $\sim 2.5$ (e.g. the stellar fraction - virial mass relation at $z \sim 4$ shown in the bottom panel of Fig.~\ref{fig:figA1} and other relations which are not shown here).

\bsp	
\label{lastpage}
\end{document}